\documentclass[reprint, superscriptaddress, amsmath, amssymb, aps, prb]{revtex4-2}

\usepackage{graphicx}
\usepackage{dcolumn}
\usepackage{bm}
\usepackage{gensymb}
\usepackage{siunitx}
\usepackage{color}
\usepackage{multirow}
\usepackage{array}
\usepackage{tabularx}
\usepackage{amsmath}
\usepackage{amssymb}
\usepackage{pifont}
\usepackage{cleveref}
\usepackage{babel}
\usepackage{subfigure}
\usepackage[version=4]{mhchem}
\usepackage{booktabs}

\usepackage{chemformula} 
\usepackage[T1]{fontenc} 
\usepackage{xcolor}
\usepackage{gensymb}
\usepackage{yfonts}

\usepackage{calligra}
\usepackage[T1]{fontenc}

\makeatletter
\renewcommand*\env@matrix[1][*\c@MaxMatrixCols c]{%
  \hskip -\arraycolsep
  \let\@ifnextchar\new@ifnextchar
  \array{#1}}
\makeatother
\newcommand{\Citation}[1]{\textcolor{red}{{\footnotesize \tt Citation required}}}

\DeclareMathAlphabet{\mathcal}{OMS}{cmsy}{m}{n}

\begin{document}

\title[]{First-principles Newns-Anderson Hamiltonian Construction for Chemisorbed Hydrogen at Metal Surfaces}

\author{Nils Hertl} 
\affiliation{Department of Chemistry, University of Warwick, Coventry, CV4 7AL, United Kingdom}
\affiliation{Department of Physics, University of Warwick, Coventry, CV4 7AL, United Kingdom}

\author{Zsuzsanna Koczor-Benda}
\affiliation{Department of Chemistry, University of Warwick, Coventry, CV4 7AL, United Kingdom}

\author{Reinhard J. Maurer}
\email{reinhard.maurer@univie.ac.at}
\affiliation{Department of Chemistry, University of Warwick, Coventry, CV4 7AL, United Kingdom}
\affiliation{Department of Physics, University of Warwick, Coventry, CV4 7AL, United Kingdom}
\affiliation{Faculty of Physics, University of Vienna,  Kolingasse 14-16, 1090, Vienna, Austria}



\begin{abstract}
The Newns-Anderson Hamiltonian is widely used to describe adsorption at gas-solid interfaces, yet its construction typically relies on simplifying assumptions such as constant coupling and the wideband limit approximation. Here, we present a first-principles approach to construct Newns-Anderson Hamiltonians by applying projection operator diabatisation to Hamiltonian matrices obtained from Kohn-Sham density functional theory calculations. We demonstrate this method for chemisorbed hydrogen on three fcc metal(111) surfaces: Al, Cu, and Pt. To validate the electronic coupling between adsorbed hydrogen and the metal surface, we compute the projected density of states, electronic tunnelling lifetimes, and vibrational lifetimes from the constructed Newns-Anderson Hamiltonians and find good agreement with reference calculations. Analysis of the chemisorption function reveals that the wideband limit approximation is valid for H/Al(111) but has limited applicability for H/Cu(111) and H/Pt(111).
\end{abstract}

\maketitle

\clearpage

\section{Introduction}
In the field of gas-surface science, the Newns-Anderson model\cite{anderson1961, newns1969} is leveraged to obtain a rationale of the  multifaceted interactions between the electronic states of gaseous surfactants and metallic surfaces \cite{yoshimori1986, gross2009}. In this model, a single electronic state of an adsorbate with energy $\varepsilon_\text{ad}$ couples to a manifold of electronic states of the metal surface. The coupling strength between the adsorbate state and a respective metal state $|s\rangle$ is characterised by a scalar $V_{s}$.
The Newns-Anderson Hamiltonian (NAH), constituted by the aforementioned quantities, can thus be utilised to characterise, for example, line shapes in scanning tunnelling spectroscopy \cite{kroger2005, lorente2000}, charge transfer \cite{nakanishi1988, langreth1991, marston1993} and electronically non-adiabatic effects \cite{brandbyge_1995_electronically, jin_practical_2019} in atom-surface interactions. 
Furthermore, the Newns-Anderson model forms the basis for the d-band model, which rationalises how surface reactivity in heterogeneous catalysis arises from the surface and adsorbate electronic structure. \cite{holloway1984, hammer1995, norskov2011,pettersson2014}
Aside from general trends that can be inferred from the Newns-Anderson model, the NAH can be leveraged to simulate chemical dynamics at surfaces\cite{yoshimori1986, gross2009}. In previous work, the model was used for numerically exact quantum dynamics with the hierarchical equation of motion (HEOM) approach \cite{tanimura1989}, which accounts for non-adiabatic and nuclear quantum effects \cite{tanimura2020}. Alternatively, NAH-based mixed quantum-classical dynamics (MQCD) simulation can be performed in which the electrons and nuclei are treated quantum mechanically and classically, respectively. Examples of such MQCD methods include independent electron surface hopping \cite{shenvi09, gardner2023} and Ehrenfest Dynamics \cite{li2005}.

The aforementioned examples underpin the importance of well-parametrised NAHs, in particular the importance of accurate couplings between the electronic states of the adsorbate and the metal. In the NAH, adsorbate and substrate states are strictly orthogonal, which provides an intuitive diabatic representation to study molecule-surface charge transfer and non-adiabatic effects in dynamics at metal surfaces. Yet, the acquisition of accurate diabatic potential energy surfaces (PESs) is not straightforward from electronic structure theory, particularly as diabatic representations are not unique. In the case of simple, low-dimensional model Hamiltonians, diabatic PESs and the molecule-metal couplings can be represented by analytical functions and often parameterised with the help of experimental quantities \cite{yoshimori1986, brandbyge_1995_electronically, erpenbeck2018, gardner2023, gardner2023a}. For the construction of high-dimensional NAHs, one has to rely on electronic structure theory, such as density functional theory (DFT). However, electronic structure methods always provide results in the hybridised, canonical basis of the eigenstates that diagonalises the many-body Hamiltonian, and thus the resulting electronic states are adiabatic. To bring \textit{ab-initio} electronic structure results into a form that is compatible with the NAH, a diabatisation procedure is required.

Commonly, NAHs are defined in the wideband limit, i.e., it is assumed that the molecule-metal coupling is energy-independent. This assumption typically holds for sp-bands \cite{gross2009}. For metals with d-bands, on the other hand, the validity of this assumption is not a given. In the wideband limit, models with two electronic states, one representing the occupied and one representing the unoccupied adsorbate state, can be parametrised directly by evaluating the diabatic adsorbate states with methods such as constrained DFT \cite{wu2005, wu2006, meng2022}  or ground-state DFT where an electronic field is employed.\cite{roy2009} If there are only two possible states and the adiabatic ground state of the system is known from a plain DFT calculation, the coupling of the system can be inferred. This approach has been successfully used to parametrise high-dimensional hamiltonians to perform independent electron surface hopping simulations of nonadiabatic effects during quantum-state-resolved molecular scattering at metal surfaces \cite{meng2024, meng2025}. Unfortunately, the results of cDFT calculations  often sensitively depend on the nature of the employed constraint \cite{kaduk2012} and this approach cannot be easily adapted to NAHs beyond the wideband limit. 


An alternative approach is the use of diabatisation and projection schemes to transform Kohn-Sham (KS) Hamiltonians from \textit{ab-initio} calculations in a post-processing fashion to obtain diabatic states and couplings that can be interpreted in the context of the NAH. These methods typically rely on partitioning the obtained KS Hamiltonian into different subsets for the sake of acquiring the couplings between the states of the individual subsets, e.g., electron donor and acceptor states. Prominent approaches are the dimer-projection method \cite{baumeier2010},  fragment-orbital DFT \cite{senthilkumar2003, valeev2006, schober2016}, or the projection-diabatisation approach \cite{kondov2007, futera_electronic_2017,  ghan2020improved, bahlke_2021_local}. The dimer-projection method generates diabatic states by combining orbitals from isolated fragments, providing a simple and computationally inexpensive route to obtain electronic couplings. While attractive for weakly interacting systems, its reliability diminishes at metal surfaces where strong hybridisation and screening may occur. The fragment-orbital DFT approach refines this idea by constructing diabatic states from fragment orbitals that are results of independent calculations on the fragments, improving charge localisation and transferability. The diabatic couplings are then approximated by computing the expectation value of the Hamiltonian operator of the combined system. Yet, the couplings depend on the choice of the fragment construction \cite{schober2016}, which can limit the accuracy of the fragment-orbital DFT approach for gas–surface interactions. The projection-operator diabatisation (POD) approach makes use of the entire KS Hamiltonian as a starting point for the partitioning \cite{kondov2007} into subsets and thus overcomes some of the limitations of the dimer-projection method and fragment-orbital DFT approach. The partitioned Hamiltonian is subsequently transformed into a block-diagonal form, where each block represents a fragment. In previous studies, the main purpose of POD was to acquire accurate couplings, as they can be interpreted as electron transfer rates between adsorbate and substrate within Marcus theory \cite{marcus1965,futera_electronic_2017, ghan2023interpreting}. Variants of the methodology, such as POD2, and in particular POD2GS, have been specifically developed to increase the accuracy of couplings for molecule-surface systems \cite{ghan2020improved}. These couplings can also be used to study the hybridisation of the orbitals relevant for the Kondo effect in molecule-surface systems \cite{bahlke_2021_local}.  However, the POD approach has thus far mostly been applied to study relatively weakly coupled systems, whereas many systems feature intermittent strong hybridisation during the dynamics.

In this work, we present a systematic study of how NAHs of strongly chemisorbed adsorbates, such as hydrogen atoms on metal surfaces, can be constructed from KS DFT Hamiltonians optimised via first-principles calculations with the goal to create an NAH that provides an accurate representation of the density-of-states, the electronic tunnelling lifetimes \cite{ghan2023interpreting} and vibrational lifetimes due to electron-phonon coupling. We apply this approach to chemisorbed hydrogen at metal surfaces with Fermi edges that are dominated either by $s$-states (Cu), $p$-states (Al), or $d$-states (Pt). The construction is sensitive to numerical settings, in particular to the completeness of the atom-centred basis set, as larger basis sets lead to issues with the projection, such that an accurate electronic representation cannot be fully reconstructed from a subset of adsorbate states. Systematic tests on observables such as the projected density of states are used to identify viable basis sets where POD projections recover the DFT electronic structure. Once viable settings are identified, interesting trends across the different surfaces can be extracted from the projected NAHs. For example, the chemisorption function for the main group metal Al differs significantly from that of the two transition metal surfaces Cu and Pt, as it shows little energy dependence in molecule-metal couplings. Cu and Pt, however, comprise a strong energy dependence at energies close to the Fermi level and above. This highlights that the commonly applied wideband limit is already questionable for transition metals with filled $d$-shells, such as the noble metals.

\section{Theory}

\subsection{Projection-Operator Diabatisation}

The POD method starts from the Kohn-Sham Hamiltonian matrix of the full interacting system, $\mathbf{H}$, expressed in a basis of non-orthogonal atom-centred orbitals $\{\phi\}$ with a finite basis set size $N$.  We omit labels for crystal momentum, $\mathbf{k}$, in this section to simplify the notation, but all matrices ($\mathbf{H}, \mathbf{C}, \mathbf{S}$) and energies are evaluated on a dense $\mathbf{k}$-grid. The eigenvalue equation of $\mathbf{H}$ can be written in the form 
\begin{equation}
    \mathbf{H}\mathbf{C}=\mathbf{S}\mathbf{C}\mathbf{E},
\end{equation}
where $\mathbf{C}$ is the coefficient matrix and $\mathbf{S}$ is the overlap matrix. 

In the POD2 variant \cite{ghan2020improved}, matrices $\mathbf{H}$ and $\mathbf{S}$ are partitioned into adsorbate-adsorbate ($aa$), substrate-substrate ($ss$), adsorbate-substrate ($as$), and substrate-adsorbate ($sa$) blocks. This is followed by L\"owdin orthogonalisation within $\mathbf{H}_{aa}$ and $\mathbf{H}_{ss}$ blocks, transforming the matrix blocks to the orthogonal basis as
\begin{align}
    \tilde{\mathbf{H}}_{aa}&=\mathbf{S}_{aa}^{-1/2}\mathbf{H}_{aa}\mathbf{S}_{aa}^{-1/2} \\
    \tilde{\mathbf{H}}_{ss}&=\mathbf{S}_{ss}^{-1/2}\mathbf{H}_{ss}\mathbf{S}_{ss}^{-1/2}. 
\end{align}

Separate eigenvalue problems are solved for the $\tilde{\mathbf{H}}_{aa}$  and $\tilde{\mathbf{H}}_{ss}$ blocks
\begin{align}
    \tilde{\mathbf{H}}_{aa}\tilde{\mathbf{C}}_a&=\tilde{\mathbf{C}}_a \mathbf{E}_a \label{eq:eig_aa} \\
    \tilde{\mathbf{H}}_{ss}\tilde{\mathbf{C}}_s&=\tilde{\mathbf{C}}_s \mathbf{E}_s \label{eq:eig_ss}
\end{align}
and the resulting $\tilde{\mathbf{C}}_a$ and $\tilde{\mathbf{C}}_s$ are transformed back to the original non-orthogonal basis as $\mathbf{C}_a=\mathbf{S}_{aa}^{-1/2}\tilde{\mathbf{C}}_a$ and $\mathbf{C}_s=\mathbf{S}_{ss}^{-1/2}\tilde{\mathbf{C}}_s$.
 A transformation matrix of the form
 \begin{equation}
   \mathbf{C}_\mathrm{t}=\begin{bmatrix}
     \mathbf{C}_a & 0\\
     0 & \mathbf{C}_s
    \end{bmatrix}	
\end{equation}

is used to project $\mathbf{H}$ to the fragment basis, 
resulting in matrix $\bar{\mathbf{H}}^\mathrm{POD2}$ that contains the diabatic orbital energies of the fragments on its diagonal, while the $\bar{\mathbf{H}}_{as}$ and $\bar{\mathbf{H}}_{sa}$ blocks contain the couplings between each of the diabatic states localized on the adsorbate and on the substrate.


\begin{equation}
\bar{\mathbf{H}}^\mathrm{POD2}=\mathbf{C}_\mathrm{t}^\dagger \mathbf{H} \mathbf{C}_\mathrm{t}=\begin{bmatrix}[ccc|ccc]
\varepsilon_{a1}   &     0         & 0     &               &               & \\
 0              & \varepsilon_{a2} &  0    &               & \bar{\mathbf{H}}_{as}        & \\
 0              &    0          &\ddots &               &               &  \\
 \hline
                &               &       & \varepsilon_{s1} &  0            & 0 \\
                & \bar{\mathbf{H}}_{sa}        &       &  0            & \varepsilon_{s2} & 0\\
                &               &       &   0           &  0            & \ddots\\
\end{bmatrix}
\label{eq:Hdiag}
\end{equation}

The POD2 diabatic states are completely restricted to fragments, which results in a qualitatively correct behaviour of the couplings with basis set size and interfragment separation \cite{ghan2020improved}. States within each fragment block are orthogonal to each other; however, states of $a$ are in general not orthogonal to $s$, since they are expressed in different sets of bases. This typically results in the overestimation of couplings, which can be corrected by an additional L\"owdin or Gram-Schmidt orthogonalisation step, yielding the POD2L and POD2GS methods \cite{ghan2020improved}, respectively. The POD2GS variant is designed for the study of adsorbates on surfaces due to the inherent asymmetry in such systems. With the additional orthogonalisation step in POD2L and POD2GS, the states are no longer strictly localised on fragments, in contrast to POD2. The improved accuracy of POD2L and POD2GS compared to POD and POD2  has been previously demonstrated \cite{ghan2020improved} by comparing against electronic couplings calculated with higher-level wavefunction methods for the Hab11 benchmark of molecular dimers \cite{kubas2014electronic}. 

In the original implementation of POD2GS for adsorbate-surface systems, only a partial GS orthogonalisation is performed, i.e. one selected adsorbate state (i) is projected out from all substrate states (j). For achieving this, a transformation matrix is built with columns $j$ defined as
\begin{equation}
    \mathbf{C}^\mathrm{GS}_j=\frac{1}{\sqrt{1-|\mathbf{S}^\mathrm{POD2}_{j,i}|^2}} (\mathbf{v}^\mathrm{substr}_j - \mathbf{S}^\mathrm{POD2}_{i,j}\mathbf{v}^\mathrm{ads}_i  )
\end{equation}
where $\mathbf{v}_{i}$ and $\mathbf{v}_j$ are unit vectors with 1.0 at position $i$ and $j$, respectively, and the overlap of POD2 states $i$ and $j$, $\mathbf{S}_{i,j}$ is calculated as
\begin{equation}
   \mathbf{S}^\mathrm{POD2}= \mathbf{C}_\mathrm{t}^\dagger \mathbf{S} \mathbf{C}_\mathrm{t}
\end{equation}
$\bar{\mathbf{H}}^\mathrm{POD2}$ is then transformed to the GS basis as
\begin{equation}
    \bar{\mathbf{H}}^\mathrm{POD2GS}= (\mathbf{C}^\mathrm{GS})^\dagger \bar{\mathbf{H}}^\mathrm{POD2} \mathbf{C}^\mathrm{GS}
    \label{eq:HPOD2GS}
\end{equation}

In our work, we orthogonalise all adsorbate states with respect to all substrate states, and perform an additional L\"owdin orthogonalization step within the $\bar{\mathbf{H}}_{ss}$ block, resulting in $\bar{\mathbf{H}}^\mathrm{POD2GS}$ with the same block structure as eq.~\ref{eq:Hdiag}. This is required to ensure that the projected density of states obtained via a Mulliken analysis (\textit{vide infra}) of  $\bar{\mathbf{H}}^\mathrm{POD2GS}$ and the Kohn-Sham Hamiltonian matrix $\mathbf{H}$ provide the same results (Figure S1 in the Supplementary Information (SI)).

\subsection{Construction of 
Newns-Anderson Hamiltonians}
The Newns-Anderson Hamiltonian \cite{anderson1961, newns1969} discretised into $N+1$ electronic states has the following form:
\begin{align}
    \mathbf{H}_\text{NA} &=
    \begin{bmatrix}[c|cccccc]
\varepsilon_\text{ad}& V_{1} & \cdots & V_{s} & \cdots & V_{N} \\ \hline 
V_{1} & \varepsilon_1 & \cdots & 0 & \cdots & 0 \\
\vdots & \vdots & \ddots & \vdots& \vdots & \vdots \\
V_{s} & 0 & \cdots & \varepsilon_s & \cdots & 0 \\
\vdots & \vdots & \vdots & \vdots& \ddots & \vdots \\
V_{N} & 0 & \cdots & 0 & \cdots & \varepsilon_N \\
\end{bmatrix},
\label{eq:NAH}
\end{align}
where $V_s$ denotes the coupling between the affinity level of the adsorbate, $\varepsilon_\text{ad}$, and the electronic state of the metal, $|s\rangle$. By comparing Eq.\,\ref{eq:NAH} with Eq.\,\ref{eq:Hdiag} one can see that $\mathbf{H}_{\text{NA}}$ can be constructed from the block-diagonalised POD Hamiltonians, $\bar{\mathbf{H}}^\mathrm{POD2}$ or $\bar{\mathbf{H}}^\mathrm{POD2GS}$, if one entry is selected from the adsorbate sub-block and the respective off-diagonal row and column are used to populate the coupling elements $V_s$.
Throughout the remainder of the paper, we select a single adsorbate state from the $\bar{\mathbf{H}}^\mathrm{POD2GS}$, denoted in the following as $a_\text{sel}$, for the construction of Newns-Anderson Hamiltonians and subsequent calculation of the spectral properties.
\subsection{Chemisorption function}
The $\bm{k}$-resolved chemisorption function, $\Delta_{\mathbf{k}}(\epsilon)$, is calculated as a weighted density of states (WDOS) for the POD2GS states, with weights determined by the adsorbate-substrate state-to-state couplings (from off-diagonal blocks of $\bar{\mathbf{H}}^\mathrm{POD2GS}_{\bm{k}}$) between a selected adsorbate state, $a_\mathrm{sel}$, and all individual substrate states. 
\begin{equation}
\Delta_{\bm{k}}(\varepsilon) = \pi \sum_{s_i} |(\bar{\mathbf{H}}^\mathrm{POD2GS}_{\bm{k}})_{a_\mathrm{sel} s_i}|^2 \delta(\varepsilon-\varepsilon_{s_i, {\bm k}}).
\label{eq:chemisorption_fuction}
\end{equation}
The $\varepsilon_{a_\text{sel}}$ and $(\bar{\mathbf{H}}^\mathrm{POD2GS}_{\bm{k}})_{a_\mathrm{sel} s_i}$ in equation \label{eq:chemisorption_fuction} are equivalent to the affinity level $\varepsilon_\text{ad}$ and the couplings with the metal states, $V_\text{s}$, in equation \ref{eq:NAH}. The broadening of the adsorbate state due to interaction with electronic states in the metal is determined from the value of the chemisorption function at the adsorbate state energy:

\begin{equation}
    \Gamma=\frac{2}{\hbar}\Delta(\varepsilon=\varepsilon_{a_\mathrm{sel}}) 
    \label{eq:Tunelling_rate}
\end{equation}

The electron tunnelling lifetime is the inverse of the broadening:

\begin{equation}
    \tau_\text{el}= \frac{1}{\Gamma}
    \label{eq:el_lifetime}
\end{equation}

When considering the electronic lifetime of a single adsorbate state, $\varepsilon_{a_{\mathrm{sel}}}$, the chemisorption function is integrated over the Brillouin zone. 
\begin{equation}
    \Delta(\varepsilon) = \sum w_{\bm{k}} \Delta_{\bm{k}}(\varepsilon)
    \label{eq:chemisorption_fuction_integrated}
\end{equation}

\subsection{Projected density of states}
We check the validity of selecting a single POD2GS adsorbate state to describe charge transfer interactions between adsorbate and substrate by reconstructing a projected density of states (PDOS) for the adsorbate in the reduced space of all POD2GS states from the substrate and the selected state of the adsorbate (thus leaving out all other adsorbate states). For this, we diagonalise $\mathbf{H}_\text{NA}$ (which is $\bar{\mathbf{H}}^\mathrm{POD2GS}$ in the reduced space)  and use a basis set analysis for separating contributions ($M_a$) of basis functions centred on the adsorbate ($\{ a \}$).

\begin{equation}
\rho_{a, \bm{k}}(\varepsilon) = \sum_p |M_{p,a, \bm{k}}|^2 \delta(\varepsilon-\varepsilon_{i, \bm{k}})
\end{equation}
where $p$ denotes the basis functions associated with the adsorbate in the reduced space.  Furthermore, $M_{p,a,\bm{k}}$, has the following expression:

\begin{equation}
    M_{p,a, \bm{k}}= \sum_{i \in \{ a\}} |C_{p,i, \bm{k}}^{\text{tr}}|^2+ \sum_{i \in \{ a \}} \sum_{j \neq i} \bar C_{p,i, \bm{k}}^{\text{tr}} S_{ij, \bm{k}} C_{p,j, \bm{k}}^{\text{tr}},
\end{equation}
where matrix $C^{\text{tr}}$ denotes the matrix that transforms the original Kohn-Sham Hamiltonian, $\mathbf{H}$, to the diagonalised form of $\mathbf{H}_\text{NA}$. Consistent with earlier works by Ghan \textit{et al.} \cite{ghan2023interpreting} on noble gas atoms adsorbed on metal surfaces, we use the absolute values of the complex Mulliken populations of the constructed NAH as taking only the real part would yield negative PDOS values due to large negative contributions of the second term. Just as the chemisorption function $\Delta(\varepsilon)$, the PDOS is subsequently averaged over the Brillouin zone, i.e.,
\begin{equation}
    \rho_a(\varepsilon) = \sum_k w_{\bm{k}} \rho_{a, \bm{k}}(\varepsilon).
\end{equation}

\subsection{Vibrational relaxation rate}
The position-dependent POD2GS-based adsorbate energies $\varepsilon_{a_\mathrm{sel}}$ and chemisorption function can be used to calculate vibrational relaxation rates as previously proposed by Brandbyge \textit{et al.} \cite{brandbyge_1995_electronically}. For every $\bm{k}$-point, the vibrational relaxation rate along the Cartesian displacement coordinate $r_\alpha$ of the hydrogen atom is
\begin{equation}
    \eta_{\bm{k}}(\bm{r}, T) = - \frac{2 \hbar}{\pi M} \int_{-\infty}^\infty d \varepsilon \left(\frac{\partial \delta_{\bm{k}}(\varepsilon, \bm{r})}{\partial r_\alpha } \right)^2 \frac{df(\varepsilon, T)}{d\varepsilon}.
    \label{eq:Fric_BWL}
\end{equation} 
Here, $M$ is the mass of the adsorbate, i.e., the H atom, and $f(\varepsilon, T)$ is the Fermi-Dirac distribution. The factor 2 in Eq.~\ref{eq:Fric_BWL} stems from the spin degeneracy. 
The authors of Ref.~\cite{brandbyge_1995_electronically} modelled adsorbate systems where the affinity level is a two-fold degenerate $\pi^*$ orbital. In the case of atomic hydrogen, the affinity level is a non-degenerate 1$s$ state. Thus, Eq.~\ref{eq:Fric_BWL} in our work and Eq.~21 in Ref.~\cite{brandbyge_1995_electronically} differ by a factor of 2. The quantity $\delta_{\bm{k}}(\varepsilon, \bm{r})$ is the phase-shift and given by the following expression:
\begin{equation}
    \delta_{\bm{k}}(\varepsilon, \bm{r}) = \mathrm{arctan}\left(\frac{\Delta_{\bm{k}}(\varepsilon, \bm{r})}{\varepsilon_{a_\mathrm{sel}, \bm{k}}(\bm{r}) + \mathcal{H}_{\bm{k}}(\varepsilon, \bm{r}) - \varepsilon} \right)
\end{equation}
$\mathcal{H}_{\bm{k}}(\varepsilon, \bm{r}) $ represents the real part of the self-energy and is connected to the couplings $\Delta_{\bm{k}}(\varepsilon, \bm{r})$ via a Hilbert transform, i.e.,
\begin{equation}
    \mathcal{H}_{\bm{k}}(\varepsilon, \bm{r}) = \frac{\mathcal{P}}{\pi} \int_{-\infty}^\infty \frac{d\varepsilon' \Delta_{\bm{k}}(\varepsilon',\bm{r})}{\varepsilon - \varepsilon'},
    \label{eq:Shift-function}
\end{equation}
where $\mathcal{P}$ denotes the Cauchy principal value. 
One can also retain the expression for the vibrational relaxation rate in the wideband limit if one assumes that the chemisorption function in Eq.~\ref{eq:Fric_BWL} is energy independent \cite{mizielinski_electronic_2005, jin_practical_2019}, i.e., $\Delta(\varepsilon, \bm{r}) = \Delta(\bm{r})$. Since the POD2GS approach provides a spectrum of couplings, we are left with a choice for the chemisorption function, which introduces a certain flavour of arbitrariness into the wideband limit analysis.  However, given that for $T=0$\,K, the derivative of the Fermi-Dirac distribution becomes 
\begin{equation}
    -\frac{df(\varepsilon, T)}{d\varepsilon} =  \delta(\varepsilon - \varepsilon_\mathrm{F}),
    \label{eq:delta_distr}
\end{equation}

with $\varepsilon_\mathrm{F}$ being the Fermi-level, picking the couplings at the Fermi-level, $\Delta(\varepsilon_\mathrm{F}, \bm{r})$ is the most natural choice.  The expression for the vibrational relaxation rate, Eq.\,\ref{eq:Fric_BWL}, can then be  modified by  setting $\Delta(\varepsilon, \bm{r})=\Delta(\varepsilon_\text{F}, \bm{r})$ and discarding the shift-function $\mathcal{H}(\varepsilon, \bm{r})$, which results in
\begin{align}
    \eta_{\bm{k}}(\bm{r}, T) &=  -\frac{2 \pi \hbar}{ M} \int_{-\infty}^\infty d \varepsilon \frac{df(\varepsilon, T)}{d\varepsilon} \rho^2_{a_\mathrm{sel},\bm{k}}(\varepsilon,  \bm{r})  \nonumber \\
    &\times \left(\frac{\varepsilon_{a_\mathrm{sel}, \bm{k}}(\bm{r})  -\varepsilon}{\Delta_{\bm{k}}(\varepsilon_\mathrm{F}, \bm{r})} \frac{d \Delta_{\bm{k}}(\varepsilon_\mathrm{F}, \bm{r})}{d r_\alpha } - \frac{d \varepsilon_{a_\mathrm{sel}, \bm{k}}(\bm{r})}{d r_\alpha }  \right)^2
        \label{eq:Fric_WL}
\end{align}
with $\rho_{a_\mathrm{sel},\bm{k}}(\varepsilon, \bm{r})$ being the adsorbate PDOS
\begin{equation}
    \rho_{\mathrm{a},{\bm k}}(\varepsilon, \bm{r}) = \frac{1}{\pi} \frac{\Delta_{\bm{k}}(\varepsilon_\mathrm{F}, \bm{r})}{\left( \varepsilon_{a_\mathrm{sel}, \bm{k}}(\bm{r})-\varepsilon\right)^2 + \Delta_{\bm{k}}^2(\varepsilon_\mathrm{F}, \bm{r})},
\end{equation} 
which is a Lorentzian distribution centred around the energy of the diabatic adsorbate state $\varepsilon_{a_\mathrm{sel}}$. To assess the role of thermal fluctuations, we also provide the expression for $T=0$\,K, which can be obtained by inserting Eq.~\ref{eq:delta_distr} into Eq.~\ref{eq:Fric_WL} and subsequent integration:
\begin{align}
    \eta_{\bm{k}}(\bm{r}) &= \frac{2 \hbar \pi}{M}  \rho^2_{\mathrm{a_{sel}}, \bm{k}}(\varepsilon_\mathrm{F}, \bm{r})   \nonumber \\
    &\times \left[\frac{\varepsilon_{a_\mathrm{sel}, \bm{k}}(\bm{r})  -\varepsilon_\mathrm{F}}{\Delta_{\bm{k}}(\varepsilon_\mathrm{F}, \bm{r})} \frac{d \Delta_{\bm{k}}(\varepsilon_\mathrm{F}, \bm{r})}{d r_\alpha } - \frac{d \varepsilon_{a_\mathrm{sel}, \bm{k}}(\bm{r})}{d r_\alpha } \right]^2,
    \label{eq:Fric_WL_0K}
\end{align}
where $\rho_{\mathrm{a_{sel}}, \bm{k}}(\varepsilon_\mathrm{F}, \bm{r})$ is the projected density of states expression within the wideband limit at the Fermi energy.
This is the expression derived by Mizielinski \textit{et al.} for the case of two degenerate spin states \cite{mizielinski_electronic_2005}.
Independent of the precise approximation made for the chemisorption function, the vibrational lifetime due to electron-hole pair excitation, $\tau_\mathrm{vib}$, is obtained by 

\begin{equation}
  \tau_\mathrm{vib}  = \frac{1}{\eta(\bm{r}, T)},
  \label{eq:vib_lifetime}
\end{equation}
where the vibrational relaxation rate is integrated over the Brillouin zone:

\begin{equation}
    \eta(\bm{r}, T) = \sum_{\bm{k}} w_{\bm{k}} \eta_{\bm{k}}(\bm{r}, T).
\end{equation}


\section{Methodology} 
\textbf{DFT Calculation details}\\
We performed DFT calculations with the PBE functional \cite{pbe1, pbe2} for three different systems: H/Al(111), H/Cu(111) and H/Pt(111) using the FHI-aims code (211010 development version) \cite{blum_ab_2009}.  The systems were modelled as a $p(3\times3)$ slab with six layers. Periodic boundary conditions were used for all three directions, and to prevent interactions with the slab and its periodic images in the perpendicular direction, we used a vacuum slab of 50\,{\AA}. The bare slab was optimised, but the two bottom layers were kept fixed to mimic the interlayer distance in a bulk. For all three systems, we calculated one-dimensional energy curves by placing the hydrogen atom directly above a metal surface atom, i.e. the top-site, and varying its altitude.  The Brillouin zone was sampled with $12\times12\times1$ $\bm{k}$-point grid employing the sampling scheme proposed by Monkhorst and Pack \cite{monkhorst76}. We further employed an atomic ZORA relativistic correction \cite{ZORA} as well as the van der Waals correction scheme by Tkatchenko and Scheffler \cite{tkatchenko_accurate_2012}. We considered our calculations to be converged when the differences in the total energy, eigenvalues, electronic density, and forces between two optimisation steps were smaller than $10^{-6}$\,eV, $10^{-3}$\,eV, $10^{-5}$\,eV/a$_0^3$, and $10^{-4}$\,eV/{\AA}, respectively. To determine the minimum energy configuration on those one-dimensional curves, we optimised the hydrogen atom $z$-coordinate but constrained the positions of the metal atoms. For all structure relaxations, a standard 'tight' basis set numerical integration grid, as per FHI-aims basis defaults, were used. For assessing the sensitivity of the quantities, which we extract with the POD2GS approach, on the basis set, we started from a Tier2 basis set, which is standard in 'tight' settings for hydrogen and systematically reduced the basis functions on the hydrogen atom. For the vibrational lifetimes, we displaced the hydrogen atom along the surface normal direction by $\pm10^{-3}$\,{\AA} with respect to the equilibrium height. This enables us to calculate the changes in the chemisorption function $\Delta(\varepsilon)$, the selected adsorbate state $\varepsilon_{a_\mathrm{sel}}$, and the shift-function $\mathcal{H}(\varepsilon)$ required to evaluate Eq.~\ref{eq:Fric_BWL} and Eq.~\ref{eq:Fric_WL}. For the computation of the \textit{ab-initio} reference data for the vibrational lifetimes, we made use of first-order time-dependent perturbation theory to compute electronic friction tensors from Kohn-Sham density functional theory \cite{hellsing1984, maurer2016}. We refer to Ref.~\cite{Box_ab_2023} for a detailed description of the underlying formulae and implementation within FHI-aims. The delta distribution was approximated by a Gaussian distribution with a width $\sigma$ set to 0.6\,eV, consistent with previous work for hydrogen on pristine transition metal surfaces \cite{box_room_2024, stark2025}. In all evaluations of the vibrational lifetimes, i.e., POD-based and TPDT reference calculations, we set the electronic temperature in the friction expression to 300\,K. For the TDPT based electronic friction calculations, we only used a standard Tier-2 basis set with a tight integration grid, as the dependence on the basis set was shown to be small in this system.\cite{sachs2025, stark2025} 

\textbf{Implementation details}
The POD2GS analysis technique was implemented in the Julia programming language. We make use of functionalities from the ACEhamiltonians.jl package \cite{zhang2022equivariant} to extract and load in data from FHI-aims calculations, such as real-space Hamiltonian and overlap matrices, and convert them to complex k-point dependent matrices. We use the efficient HDF5 format for transferring data.

We use the full adiabatic Hamiltonian to construct POD2GS states, in contrast to Ghan et al., who replaced the substrate-substrate block with the adiabatic Hamiltonian of a pristine substrate without adsorbate \cite{ghan2023interpreting}. While this simplification might work for physisorbed systems, in the case of chemisorption, the substrate block is expected to be significantly altered by the strong interaction between adsorbate and substrate.

\textbf{State following}. We introduce functionality to automatically select the POD2 adsorbate state ($a_\mathrm{sel}$) that has maximal similarity to a reference adsorbate state. 
\begin{equation}
a_\mathrm{sel} = \text{argmax}_a(|\Re\sum_{i \in \{ \mu \}} \sum_{j}  ({c}^\dagger_\mathrm{ref})_i S_{ij} (C_t)_{a,j})|)
\end{equation}
where $c_\mathrm{ref}$ contains the coefficients of basis functions centred on the adsorbate ($\{ \mu \}$) for the reference adsorbate state. 
This reference can be, for example, a molecular orbital from a calculation on the isolated adsorbate or a POD2GS adsorbate state from a different geometry. By updating the reference state at each geometry of a distance scan, for example, this enables a maximum overlap-like state-following, which accounts for smooth changes in the character of the state with varying distance. We found that state-following was necessary to yield smooth properties along binding curves for hydrogen basis sets with a large number of basis functions{---}see Fig.S2 SI.
Thus, for calculations with basis sets containing multiple groups of basis functions with the same angular momentum character, we selected the initial reference as the first occupied (for spin major) and unoccupied (for spin minor) POD2GS adsorbate state when the adsorbate is 6\,{\AA} from the surface, and followed this state during the atom-surface distance scan.

\section{Results}
\begin{figure}
    \centering
    \includegraphics[width=3.3in]{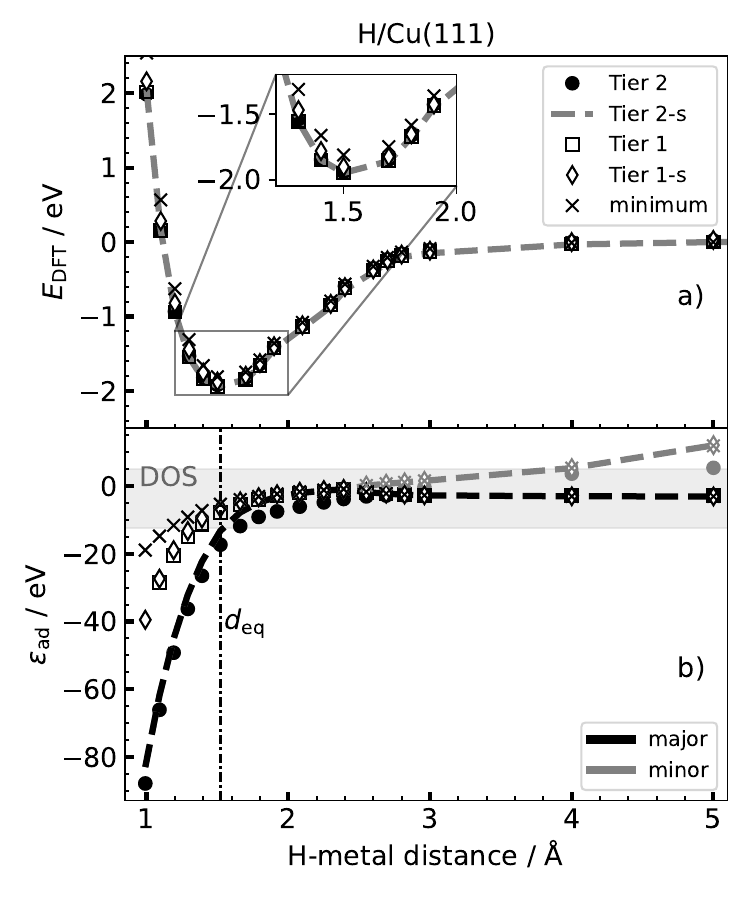}
    \caption{Ground state DFT energy, $E_\text{DFT}$, and lowest diabatic adsorbate states, $\varepsilon_\mathrm{ad}$, obtained with the POD approach as a function of the H-Cu distance and NAO basis set plotted in panel a) and b), respectively. The energies in panel a) are referenced against the interaction energy for hydrogen placed 5\,{\AA} above the surface, whereas the $\varepsilon_\mathrm{ad}$ values are referenced against the Fermi energy. The grey-shaded region indicates the energy range of the valence electron density of states of Cu(111). The vertical dashed line indicates the equilibrium distance $d_\mathrm{eq}$ of the hydrogen atom when adsorbed at the top-site. The grey and black colour in panel b) indicates the spin state that is predominantly occupied (\textit{major}) or unoccupied (\textit{minor}), respectively. }
    \label{fig:ead_HCu}
\end{figure}

\subsection{Benchmark of the POD-based quantities}

As the POD method relies on an atomic orbital-based partitioning of the Hamiltonian, the derived quantities can become unreliable when there is a significant overlap and hybridisation between adsorbate and substrate atomic orbitals, which is dependent on the adsorbate-substrate distance and the parameters of the applied basis functions. We start with the investigation of how strongly the POD-based quantities are affected by the basis set utilised in the DFT calculations. To do so, we pick one system, H/Cu(111), and systematically reduce the number of basis function of the standard 'tight' settings in FHI-aims on hydrogen. FHI-aims uses numerical atomic orbitals (NAOs) as basis functions.\,\cite{blum_ab_2009} For hydrogen, the default 'tight' settings include elemental valence states and additional Tier1 and Tier2 basis functions. We do not modify the basis functions of copper because we are interested in constructing Hamiltonians that represent the adsorbate as a single-particle state interacting with a manifold of substrate states. DFT calculations were performed where we systematically removed the basis functions constituting the Tier2 and the Tier1 basis. Higher principal number s orbitals mix into the POD state at small intersystem distances, which affects the results considerably. To test the role of these s basis functions, we ran tests where we systematically removed the s-orbitals in the Tier1 and Tier2 basis sets, which we henceforth label Tier1-s and Tier2-s basis sets, respectively. Due to the mixing of the states with the same angular quantum numbers, we made use of the state following approach for the calculations with the Tier1, Tier2-s, and Tier2 basis sets. Finally, we also used a basis set that only contains the H atom 1s orbital, which we will refer to as the minimum basis. 

Figure\,\ref{fig:ead_HCu} shows the adiabatic ground state energies and the lowest diabatic adsorbate energies for an H atom positioned above a Cu atom as a function of distance. Far away from the surface, i.e. 5\,{\AA}, one of the H atom spin states is occupied (major spin state), whereas the other one is unoccupied (minor spin state). In the interaction-free region, these two spin states are energetically well separated because of the Coulomb interaction, and this energy difference is more pronounced for smaller basis sets. When the H atom approaches the surface, the two spin states start to hybridise with the electronic states of the metal surface. This interaction leads to the reduction of the energy difference between the two spin states until both states are completely degenerate, which happens between 2\,{\AA} and 3\,{\AA}, depending on the chosen basis set. The energy of the diabatic adsorbate states decreases rapidly from this point on for all basis functions, but the shifts towards lower energies are more pronounced for basis sets with larger basis functions. This results in diabatic state energies which are far away from the spectral range of the metal surface conduction band. In the case of the Tier2 basis, this occurs already at the equilibrium adsorbate-substrate distance. We find the same trend for H/Al(111) and H/Pt(111) as well (see Fig.~S3 in the SI).  

\begin{figure}
    \centering
    \includegraphics[width=3.3in]{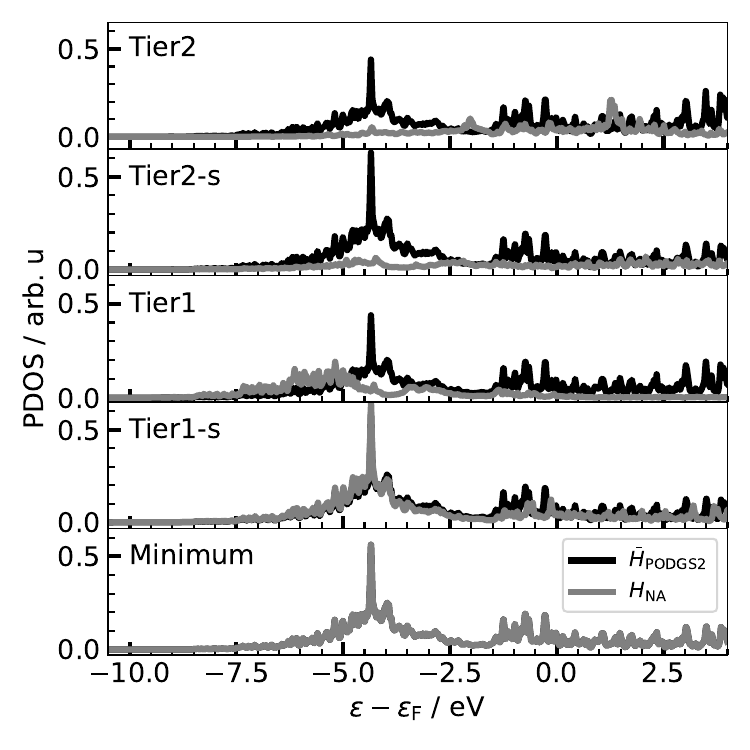}
    \caption{ POD-based PDOS of hydrogen in H/Cu(111) for various NAO basis sets specified in the top left corner of the individual panels. The grey lines show the result using only the adsorbate's selected diabatic state. The black lines represent the H atom PDOS incorporating all diabatic adsorbate states.}
    \label{fig:PDOS_HCu_tiers}
\end{figure}

Next, we assess the validity of the POD approach by investigating how well we can reproduce the hydrogen atom PDOS when using all substrate diabatic states but only the selected diabatic state of the adsorbate. Recall that the PDOS from the complete HPOD2GS Hamiltonian corresponds to the PDOS one would obtain from a Mulliken analysis applied to the Kohn-Sham Hamiltonian, $H$. The comparison between the PDOS from the Newns-Anderson Hamiltonian, $\mathbf{H}_\text{NA}$ (see Eq.~\ref{eq:NAH}) and the PDOS obtained from $\bar{\mathbf{H}}^{\mathrm{POD2GS}}$ ( Eq.~\ref{eq:HPOD2GS}) in Figure\,\ref{fig:PDOS_HCu_tiers} for the different FHI-aims basis sets shows that the agreement between the the two gets worse with increasing number of  adsorbate basis functions. 
We obtain a perfect agreement for the minimum basis set, which is not surprising given that this basis set contains a single 1s basis function and the NAH construction occurs without information loss. Hence, the space is not reduced, and both curves are exactly the same. Only in the case of the Tier1-s basis set, the shape of the reconstructed PDOS manages to reproduce the hydrogen atom PDOS constructed from all diabatic states. The strong basis-set dependence observed in the PDOS obtained from diagonalising $\mathbf{H}_\text{NA}$ arises because the NAH construction represents the adsorbate by a single effective state. This approximation is justified when the adsorbate state is already well represented by a single basis state in the original basis. However, when the basis set is larger, the adsorbate subspace becomes larger in dimension and mnay basis states contribute to the hybridised canonical eigenstates. Projecting this multi-dimensional subspace onto a single orthogonalised state introduces a truncation that depends sensitively on the basis and orthogonalisation procedure. In such cases, selection of a single state leads to information loss and the DFT PDOS cannot be fully recovered by the reduced-dimensional NAH.

Based on Figure\,\ref{fig:PDOS_HCu_tiers}, one could conclude that it would be best to continue with the minimum basis set. Yet, the minimum basis set is not sufficiently converged as it shows deviations in the depth and curvature around the minimum of the one-dimensional energy curve (Figure\,\ref{fig:ead_HCu}a). 

Another way to assess the reliability of the calculated couplings between the selected adsorbate state and the continuum of metal states, which we attained from our POD approach, is to analyse the electron tunnelling rates, $\Gamma$, and their relation to the Brillouin zone integrated chemisorption function (Eq.~\ref{eq:Tunelling_rate}). The inverse value of the electronic tunnelling rate can be interpreted as the lifetime of a hole or an electron in the 1$s$ spin orbital, respectively (Eq.~\ref{eq:el_lifetime}). The electronic tunnelling lifetimes for H/Cu(111) at the hydrogen atom equilibrium height computed with the five different basis sets are given in Table~\ref{tab:Lifetimes_HCu}. We find that an increase in the number of basis functions results in higher $\tau_\text{el}$ values. This can be explained with the diabatic adsorbate states, which are pushed to lower energies outside the range of the substrate DOS by the orthogonalisation procedure in the construction of $\bar{\mathbf{H}}^\text{PODGS2}$. The diabatic energy curves for the different adsorbate basis sets are shown in Figure~\ref{fig:ead_HCu}b). Mizielinski \textit{et al.} reported electronic tunnelling rates for H/Cu(111), albeit by fitting a Lorentzian to the PDOS of the H atom 1$s$ orbital, which is only strictly valid in the wideband limit \cite{mizielinski_electronic_2005} and therefore not directly comparable to our electronic tunnelling lifetimes. Yet, their reported value of 0.14\,fs  for hydrogen at the equilibrium distance is consistent with values we obtain with the three smallest basis sets. 

As a final benchmark, we compute vibrational lifetimes of the perpendicular hydrogen-metal stretch mode $\tau_\text{vib}$ due to electron-phonon coupling with the quantities obtained from our POD workflow (Eq.~\ref{eq:Fric_BWL}) for the five different basis sets. From the results (Table~\ref{tab:Lifetimes_HCu}), one can infer that the values for $\tau_\text{vib}$ decrease with the size of the hydrogen atom basis functions. This is connected with the rate of change of $\varepsilon_\text{ad}$ during perpendicular displacement of the hydrogen atom: the more basis functions are located on the hydrogen atom, the higher is the derivative of $\varepsilon_\text{ad}$ with respect to the  $z$-coordinate of the hydrogen atom (Fig.~\ref{fig:ead_HCu}b).  A larger spatial derivative of $\varepsilon_\text{ad}$ will, in turn, result in higher vibrational relaxation rates and consequently lead to smaller vibrational lifetimes. Unfortunately, to the best of our knowledge, there are no experimental values for the vibrational lifetimes of the perpendicular stretch mode of an H atom chemisorbed on Cu(111) at $\theta=0.11$\,ML coverage. Instead, we employ first-order perturbation theory at the level of density functional theory to compute reference values from first-principles for a thorough assessment of our attained POD-based vibrational lifetimes \cite{maurer2016, Box_ab_2023}. 

In conclusion, we find that the observables obtained from the POD approach show a strong basis set dependence. When the number of basis functions in the original electronic structure calculation is large and the basis set involves diffuse functions, the reduced-dimensional NAH constructed with the POD approach is not capable of recovering the spectral properties. Mitigating this basis set dependence likely will require the development of custom basis sets. For now, we identified the Tier1-s basis set to be the best compromise between accurately describing converged ground-state potential energy curves and providing POD-based diabatic adsorbate energies and couplings which are able to accurately reproduce observables such as the hydrogen atom 1$s$ PDOS and $\tau_\text{vib}$. We find the same behaviour for H/Al(111) and H/Pt(111) (see Fig.~S4 and Table I in the SI). In the remainder of this work, we show results based on DFT calculations with a  Tier1-s NAO basis set for hydrogen unless stated otherwise.

\begin{table}[]
    \centering
    \begin{tabular}{l c c c c c c c} \hline
          &      &         & \multicolumn{2}{l}{Basis set type} &       &            &      \\ \hline
          & min.  & Tier1-s & Tier1           & Tier2-s          & Tier2 & Lit. & TDPT \\ \hline
$\tau_\text{el}$ / ps  & 0.3  & 0.37    & 0.29            & 3.3              & 8.92  & 0.14 \cite{mizielinski_electronic_2005}       & \ding{55}    \\
$\tau_\text{vib}$ / ps & 1.46 & 1.58    & 1.93            & 0.14             & 0.1   & \ding{55}          & 1.73 \\ \hline
    \end{tabular}
    \caption{POD-based electronic tunnelling lifetimes, $\tau_\text{el} $, and vibrational lifetimes, $\tau_\text{vib}$, of H/Cu(111) for various NAO basis sets starting from the minimum basis set (labelled 'min' in the table) to the Tier2 basis set used in standard 'tight' settings. Literature values, labelled as 'Lit.', for the electronic tunnelling lifetimes are reported by Mizielinski \textit{et al}.~\cite{mizielinski_electronic_2005}. Furthermore, we conducted DFT-based time-dependent perturbation theory calculations to acquire reference values for the vibrational lifetimes. The electronic temperature in all vibrational lifetime calculations was set to 300\,K.}
    \label{tab:Lifetimes_HCu}
\end{table}

\subsection{Diabatic states and couplings for H/metal(111) systems}

\begin{figure}
    \centering
    \includegraphics[width=3.3in]{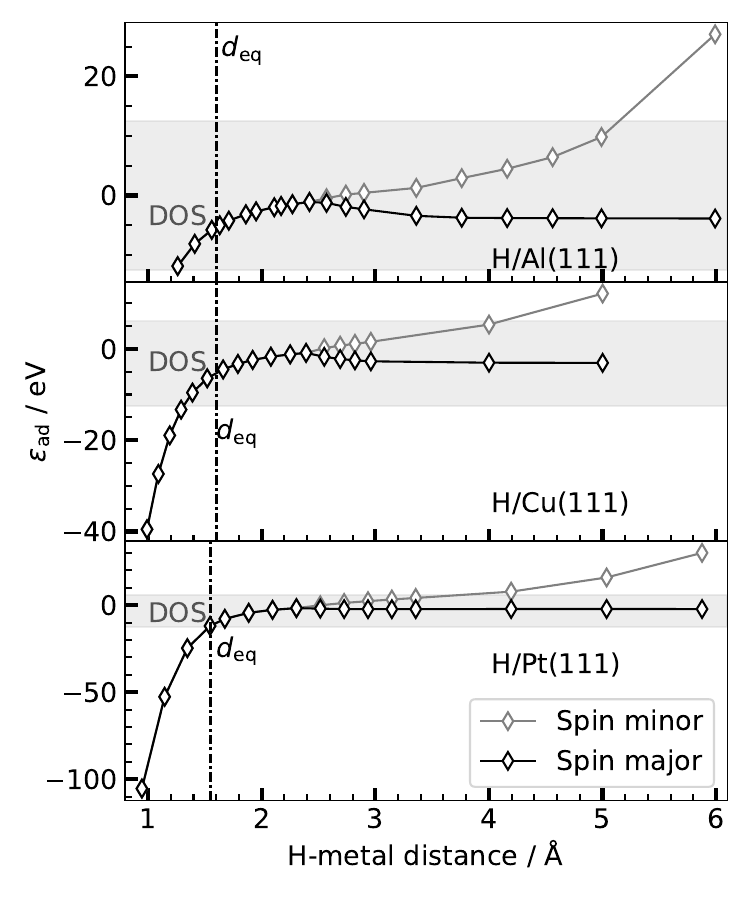}
    \caption{Lowest diabatic adsorbate state energies $\varepsilon_\mathrm{ad}$ for three different fcc metals obtained with the POD approach as a function of the H-metal(111) distance. The grey-shaded region indicates the energy range of the valence electron density of states of the respective substrate. The vertical dashed line indicates the equilibrium distance $d_\mathrm{eq}$ of the hydrogen atom when adsorbed on the top-site. The labels \textit{major} and \textit{minor} indicate the spin state that is predominantly occupied or unoccupied, respectively. }
    \label{fig:Adsorbate_energies_Hmetals}
\end{figure}

\begin{figure}
    \centering
    \includegraphics[width=3.3in]{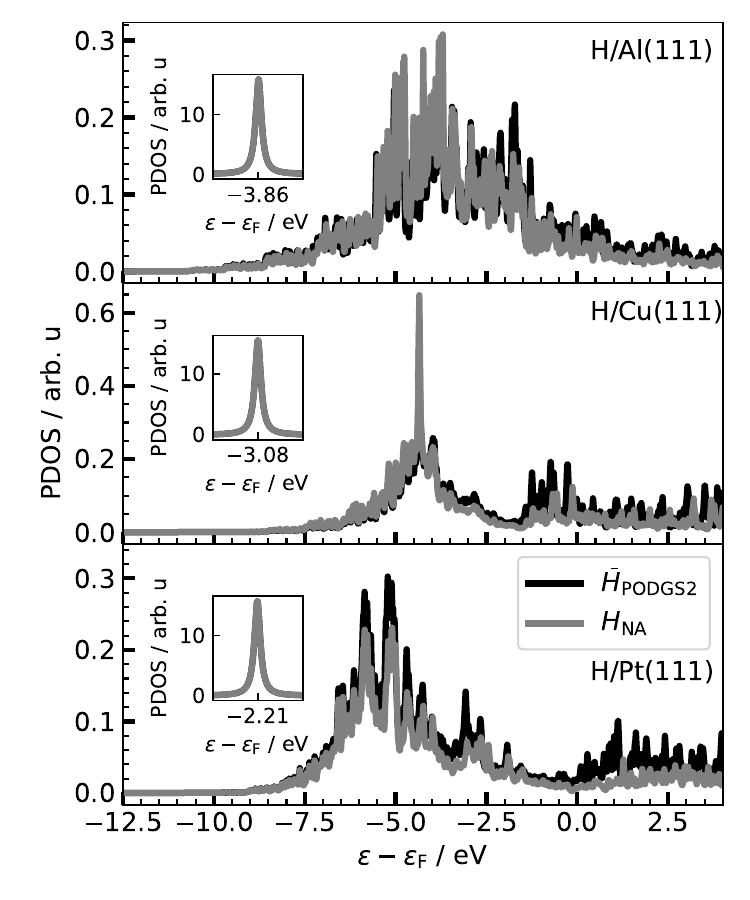}
    \caption{POD-based PDOS of hydrogen chemisorbed on different fcc metal(111) surfaces specified in the individual panels. The grey lines show the result using only the selected adsorbate diabatic state. The black lines represent the H atom PDOS incorporating all diabatic adsorbate states. The inset, with a width of 0.5\,eV, in each panel shows the PDOS of the major spin state for hydrogen in the interaction-free region.}
    \label{fig:PDOS_Hmetals}
\end{figure}

Thus far, we have studied H/Cu(111), which is a system where the adsorbate states and the conduction band have predominantly $s$-character. Now we demonstrate that our POD approach can also be applied to other metals whose conduction bands are dominated either by $p$-states, such as aluminium, or $d$-states such as platinum, where the adsorption and thus the hybridisation is stronger.  Figure\,\ref{fig:Adsorbate_energies_Hmetals} compares the diabatic adsorbate state energies for H/Al(111), H/Cu(111) and H/Pt(111). Like for H/Cu(111), the H atom is placed directly above one of the metal atoms forming the surface. The shape of the three sets of diabatic curves is qualitatively similar. For all three systems, the two spin states become degenerate around $\sim$2.5\,{\AA} which is in line with previous observations \cite{mizielinski_electronic_2005, lindenblatt_ab_2006-1, box_room_2024}. Yet, when the H atom gets past the equilibrium height, the drop in the diabatic energies becomes more pronounced in the order H/Al(111) < H/Cu(111) < H/Pt(111). This can be connected to the larger overlap of adsorbate and substrate basis functions, e.g. Pt has significantly more diffuse basis functions that have a large overlap with H basis functions already at larger separations. As a result, the orthogonalisation leads to a large shift in the adsorbate state.
Figure\,\ref{fig:PDOS_Hmetals} shows the PDOS for hydrogen at its equilibrium height at the top site of the three different fcc(111) surfaces. The insets in Figure\,\ref{fig:PDOS_Hmetals} show the PDOS for hydrogen in the interaction-free region as a reference point.  There, the PDOS is a Lorentzian with a narrow broadening that arises from the numerical evaluation. Despite the different shapes of the PDOS of the chemisorbed H atom, our POD approach manages to reconstruct the adsorbate PDOS using only one POD diabatic state.  The overall good agreement gives us confidence that our overlap matrices, and therefore the couplings, i.e., the off-diagonal elements of $\bar{\mathbf{H}}^\mathrm{POD2GS}$, which we obtain through our POD scheme, are reliable and physically meaningful. Furthermore, it demonstrates that our POD approach is robust across different metal substrates and thus broadly applicable for hydrogen-metal interactions.

\begin{figure}
    \centering
    \includegraphics[width=3.3in]{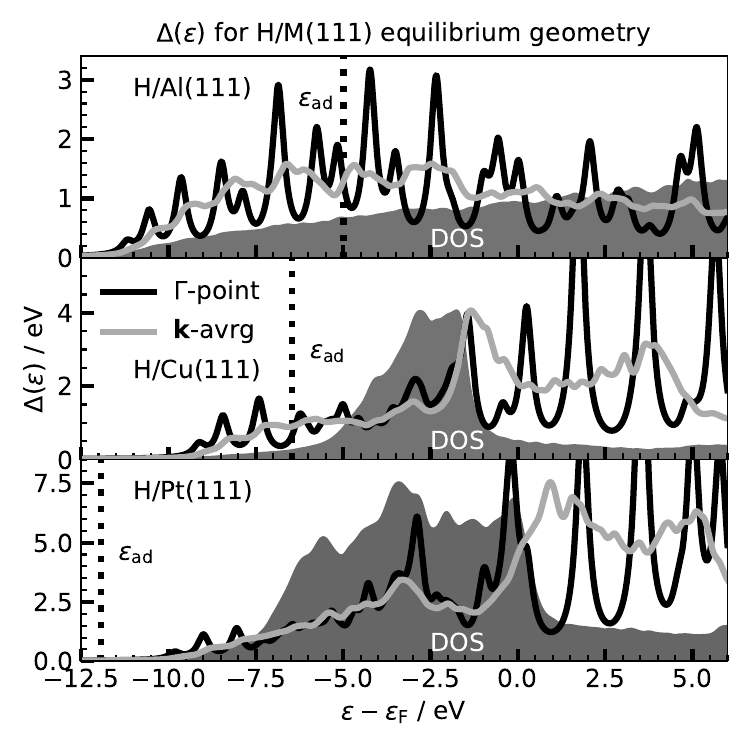}
    \caption{POD-based chemisorption function, also called WDOS, at the $\Gamma$-point (black line) and $\bm{k}$-averaged chemisorption function (grey line) for the three H/M(111) systems at their equilibrium height on the top-site. The horizontal black, dashed lines indicate the energy of the diabatic ground state $\varepsilon_\mathrm{ad}$. }
    \label{fig:WDOS_Hmetals}
\end{figure}

By comparing the chemisorption function for hydrogen at its equilibrium adsorption height across the three different substrates, the differences in adsorbate-metal interactions become apparent. Chemisorption functions obtained with Equation~\ref{eq:chemisorption_fuction} for the $\Gamma$-point and Equation~\ref{eq:chemisorption_fuction_integrated} averaged over all  $\bm k$-points differ significantly (Figure\,\ref{fig:WDOS_Hmetals}). Both differ from the density of states of the metal in highly nontrivial ways.  At the $\Gamma$-point, the chemisorption function exhibits strong fluctuations in magnitude over that energy range, as it is dominated by few states. On the other hand,  the $\bm{k}$-averaged chemisorption function is much smoother. This indicates a pronounced $\bm k$-point dependence of the chemisorption function.  Interestingly, the $\Gamma$-point and the $\bm{k}$-averaged chemisorption functions for H/Cu(111) and H/Pt(111) are similar on a qualitative level: they continuously rise from -12.5\,eV to $\sim$-3.0\,eV, and then exhibit changes in magnitude over several eVs in the energy range from about -3\,eV to 6\,eV. This is in stark contrast to H/Al(111), where, over the entire energy range of the substrate DOS, the $\bm{k}$-averaged  chemisorption function stays nearly constant. The dashed lines in Figure\,\ref{fig:WDOS_Hmetals} represent the selected adsorbate state, and with increasing atomic number of the metal, the adsorbate state is pushed to the lower edge of the energy range of the substrate density of states. This is again a consequence of strong hybridisation and the shift induced by orthogonalisation of adsorbate and substrate states.

\subsection{The Validity of the wideband limit approximation}
\begin{figure}
    \centering
    \includegraphics[width=3.3in]{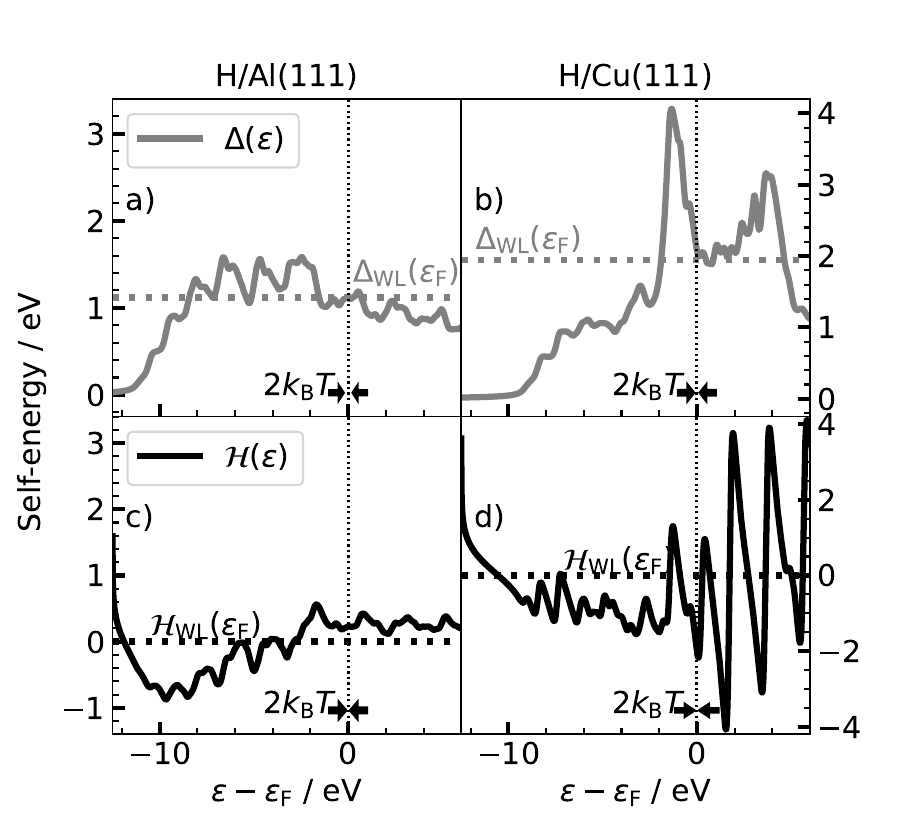}
    \caption{Imaginary part $\Delta(\varepsilon)$ and real part $\mathcal{H}(\varepsilon)$ of the self-energy obtained from the POD approach for H/Al(111) (panel a and panel c) and H/Cu(111) (panel b and d). The grey horizontal dotted lines indicate the chemisorption function under the wideband limit approximation (constant coupling) with a coupling strength of $\Delta(\varepsilon_\text{F})$.
    The black dotted horizontal line corresponds to the shift function, $\mathcal{H}(\varepsilon_\text{F})$ under the wideband limit.  The thin black dotted line represents the energy range of $2k_\text{B}T$ at 300\,K, which marks the part of the self-energy spectra that is relevant for the evaluation of the vibrational relaxation rates via Eq.~\ref{eq:Fric_BWL}.}
    \label{fig:Comparison_WL}
\end{figure}
We now discuss the calculated chemisorption functions in the context of the widely applied wideband limit approximation. We restrict the discussion to the difference between H/Al(111) and H/Cu(111) given the qualitative similarities in the energy dependence of $\Delta(\varepsilon)$ between H/Cu(111) and H/Pt(111){---}see Fig.~\ref{fig:WDOS_Hmetals} b) and c).  The wideband limit approximation assumes a constant density of states over an energy range much wider than the chemisorption function $\Delta(\varepsilon)$ \cite{yoshimori1986}. Moreover, albeit technically not the same, a constant coupling between the electronic state of the adsorbate and the electronic states of the metal is assumed, but often both approximations are invoked under the term wideband  limit approximation~\cite{mizielinski_electronic_2005, shenvi09, miao2017, miao2019, gardner2023a, meng2024}. In Figure\,\ref{fig:Comparison_WL}a) and b), we compare POD-based chemisorption functions to chemisorption functions under the wideband limit for H/Al(111) and H/Cu(111), respectively. The amplitude for the chemisorption function under the wideband limit, $\Delta_\text{WL}$, was set to the value of the POD-based chemisorption function at the Fermi-level. We chose this value as the integral for the Markovian friction coefficient in Eq.\,\ref{eq:Fric_BWL} has only a finite contribution around $\varepsilon_\text{F}$ \cite{Box_ab_2023}. In the case of H/Al(111), the wideband limit is a reasonable assumption as the POD-based chemisorption function for H/Al(111) is fairly constant from about -8\,eV to 6\,eV (see Fig.\,\ref{fig:Comparison_WL}a). For H/Cu(111), on the other hand, the wideband approximation is not consistent with the chemisorption function extracted with the POD approach, despite one would na{\"i}vely think otherwise, given that the DOS of Cu(111) around the Fermi level is dominated by $s$-states and is therefore flat. Nevertheless, $\Delta(\varepsilon)$ exhibits strong changes in magnitude from about -3\,eV to 6\,eV (see Figure\,\ref{fig:Comparison_WL}b). The pronounced energy dependence of $\Delta(\varepsilon)$ starting from $\varepsilon_\text{F}$ onwards arises from different coupling strengths $\bar{\mathbf{H}}_{as}$ for the different substrate states, {$\varepsilon_s$}, as the DOS for Cu(111) in that spectral range is fairly constant in that region. 
Below the Fermi-level, the $d$-states contribute to the DOS, which leads to a significant energy dependence. Consequently, the energy dependence in $\Delta(\varepsilon)$ arises from both changes the energy dependence of $\bar{\mathbf{H}}_{as}$ and the energy dependence of the DOS. Below the Fermi-level, where the $d$-states contribute to the DOS, we cannot make this unambiguous assignment as the DOS is not constant any longer (see Fig.\,\ref{fig:WDOS_Hmetals}b)).


Another consequence of a non-constant chemisorption function is that the shift functions $\mathcal{H}(\varepsilon)$ give finite values. The shift functions have been discussed by Newns~\cite{newns1969} for a semi-elliptic band as well as by Haldane and Anderson~\cite{haldane1976} for a magnetic impurity embedded in semiconductors. Shift functions derived from first-principles chemisorption functions are not commonly reported in literature. Figure\,\ref{fig:Comparison_WL}c) and d) show the shift functions for H/Al(111) and H/Cu(111), respectively. For H/Al(111), the shift function shows overall rather small values, consistent with the weak deviation from the wideband limit. Interestingly, $\mathcal{H}(\varepsilon)$ is negative for low-lying metal states but becomes positive close to the Fermi level. The shift function for H/Cu(111) is also negative for energy states deep below the Fermi level, but exhibits large fluctuations close to the Fermi energy. This is in stark contrast to H/Al(111), where the shift function is close to zero over a wide range of energies above and below the Fermi level.

Whether or not the wideband limit approximation is justified depends on the observable of interest. Properties that probe a narrow energy range near a specific energy, such as electron tunnelling lifetimes, are only sensitive to the chemisorption function in proximity to the adsorbate state. For the wideband limit to hold for this quantity, the chemisorption function should be constant around the adsorbate state energy.  For vibrational lifetimes due to electron-hole pair excitations, on the other hand, 
which depend on the chemisorption function near the Fermi level (cf Eq.~\ref{eq:Fric_BWL}), the wideband limit would require the chemisorption function to be constant around the Fermi level. Yet, given that the wideband limit assumes constant couplings across the entire energy range, a certain value for $\Delta(\varepsilon)$ must be picked. For example, to compute both $\tau_\text{el}$, and $\tau_\text{vib}$ within the wideband limit would therefore require $\Delta(\varepsilon_\text{ad}) = \Delta(\varepsilon_\text{F})$. This holds reasonably well for H/Al(111), but no single value can be selected for all relevant energy ranges in the cases of H/Cu(111) and H/Pt(111) (Fig.\,\ref{fig:WDOS_Hmetals} and Fig\,\ref{fig:Comparison_WL}). 


\subsection{Electron tunnelling times and vibrational lifetimes}
\begin{figure}
    \centering
    \includegraphics[width=3.3in]{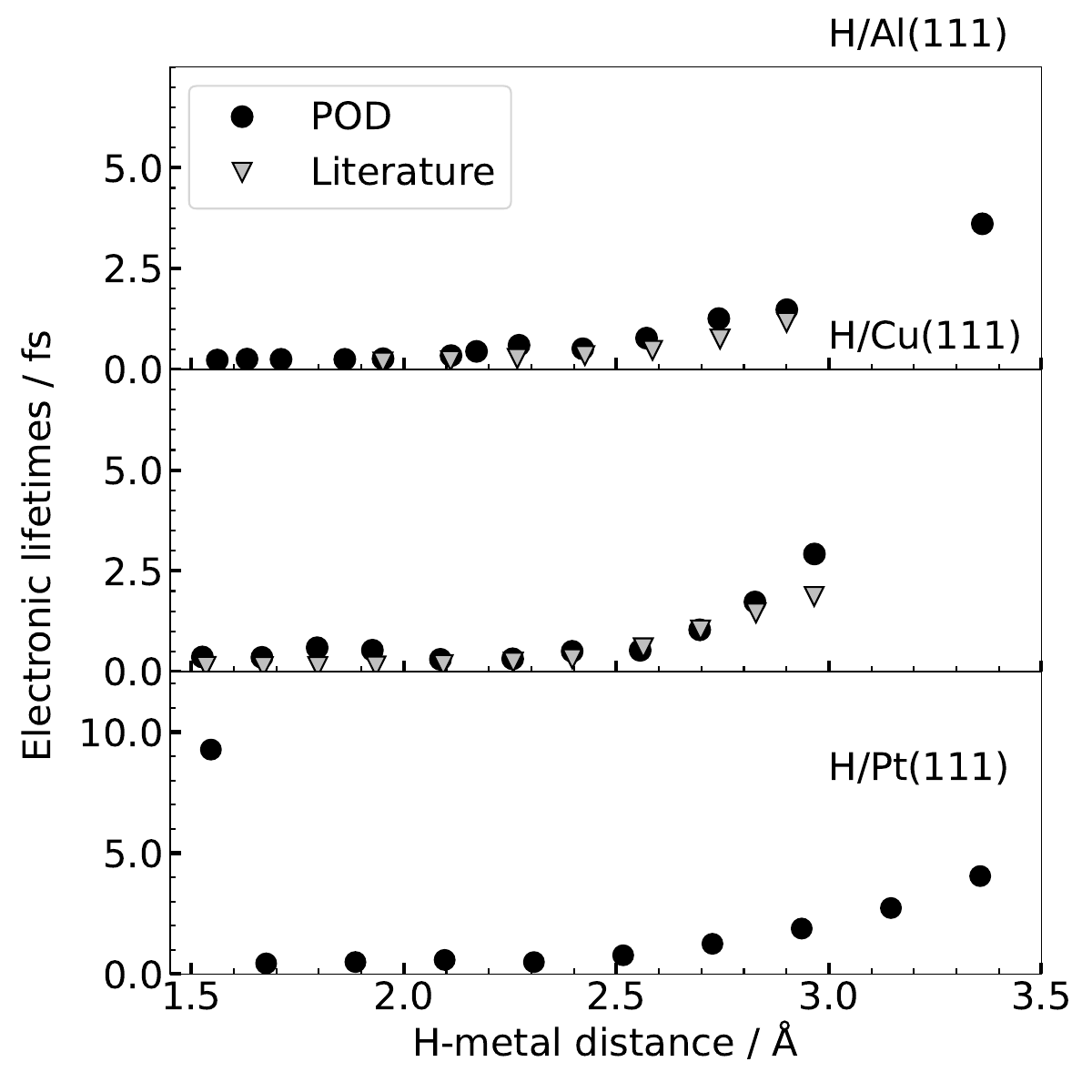}
    \caption{Electronic tunnelling lifetimes $\tau_\text{el}$ for hydrogen on the three different metal systems at various vertical H-metal distances. The values taken from the literature for comparison{---}presented here as grey markers{---}were taken from Lindenblatt \& Pehlke \cite{lindenblatt_ab_2006-1} and Mizielinski \textit{et al.} \cite{mizielinski_electronic_2005} for H/Al(111) and H/Cu(111), respectively.}
    \label{fig:Tunnelling_lifetimes_Hmetals} 
\end{figure}
With the energy-dependent chemisorption function $\Delta(\varepsilon)$ in hand, we determine the electron tunnelling lifetimes for the three H/metal systems at varying vertical distances. According to Langreth, those electronic tunnelling lifetimes are a measure for the breakdown of adiabaticity occurring in the vibrational relaxation dynamics of atoms or molecules adsorbed on surfaces \cite{langreth1985}.   The obtained lifetimes for the major spin channel can be seen in Figure\,\ref{fig:Tunnelling_lifetimes_Hmetals}.  The electronic tunnelling lifetimes show a universal behaviour for all three metal substrates. 
Far away from the surface, the lifetimes are very large due to the absence of interactions between the diabatic adsorbate state and the electronic states of the substrate. Once the hydrogen atom approaches the surface, the adsorbate state starts to hybridise with the substrate states, which leads to a broadening of the adsorbate state, and thus the electronic tunnelling lifetimes decrease. Lindenblatt \& Pehlke \cite{lindenblatt_ab_2006-1} and Mizielinski \textit{et al.} \cite{mizielinski_electronic_2005} derived the same quantity from the half width at half maximum of a Lorentzian fitted to the PDOS of the 1s orbital for H/Al(111) and H/Cu(111), respectively, which are represented by the grey markers in Figure\,\ref{fig:Tunnelling_lifetimes_Hmetals}. As one can see, both methods are in good agreement with each other.  This agreement in the case for H/Cu(111) can be explained by the fact that the chemisorption function of this system is approximately constant in the range between -7.5\,eV and -5.0\,eV and thus the wideband limit for the acquisition of $\tau_\text{el}$ can be justified if a value of the chemisorption within this energy range is selected.
Below $\sim$2.5\,{\AA}, the electronic tunnelling lifetimes for the three different systems are in the sub-fs regime except for H/Pt(111). This deviation can be explained by the location of the adsorbate state, which is located at the lower bound of the DOS of Pt(111) ( Figure\,\ref{fig:Adsorbate_energies_Hmetals}). This is a consequence of orthogonalising the adsorbate state to the substrate states. In the case of Pt(111), which has more and more diffuse basis functions in the substrate block, the mixing with the adsorbate basis functions is stronger, and consequently, orthogonalisation pushes the diabatic ground state towards lower energies. Unfortunately, we are not aware of any experimental values to which we could compare our obtained electronic tunnelling lifetimes directly.

With the help of Eq.~\ref{eq:Fric_BWL}, we determine the friction coefficient for each sampled $\bm{k}$-point in the Brillouin zone, and after subsequent integration over the Brillouin zone, we calculate the lifetime of the perpendicular stretch mode of a chemisorbed hydrogen atom via Eq.~\ref{eq:vib_lifetime}. The results are compared to vibrational lifetimes calculated from first-principles based on time-dependent perturbation theory applied to density functional theory (TDPT) \cite{maurer2016, Box_ab_2023}.  In addition, we also compute the vibrational lifetime for friction coefficients obtained within the wideband limit via Eq.~\ref{eq:Fric_WL}. All results are given in Table~\ref{tab:Vib_lifetimes}.

The vibrational lifetimes from POD with and without the wideband limit approximation agree well with the vibrational lifetime obtained from TPDT calculations   All vibrational lifetimes are in the range of a few picoseconds, and in the case of H/Cu(111), even quantitative agreement between POD and TDPT is obtained. The overestimation of the vibrational lifetime for H/Pt(111) in comparison to TDPT and the experimental lifetime can be explained by the fact that the lowest diabatic state lies energetically outside of the density of state of the Pt(111) surface (cf Figure~\ref{fig:WDOS_Hmetals}). This arises from the orthogonalisation of the adsorbate state with respect to the substrate states.  This results in weak couplings and consequently leads to larger vibrational lifetimes. When the minimal basis for hydrogen is used, the diabatic energy states is shifted to$\sim 8.6$\,eV (see Table II in the SI), which is inside the spectral range of the surface's DOS, and thus the vibrational lifetime reduces to 0.42\,ps (see Table I in the SI). This shows that for the Pt surface, the basis set dependence leads to deviations in vibrational lifetimes that are hard to control.  The differences between the values for the vibrational lifetimes within and beyond the wideband limit can be mainly attributed to the absence of the shift-function, and its spatial derivative in Eq.\,\ref{eq:Fric_WL}.
\begin{table}[]
    \centering
    \caption{Vibrational lifetimes $\tau_\text{vib}$ acquired through different methods: the equation proposed by Brandbyge \textit{et al.}\cite{brandbyge_1995_electronically} (Eq.\,\ref{eq:Fric_BWL}), the friction expression in the wideband limit (WL) with finite temperature (Eq.\,\ref{eq:Fric_WL}) and at 0\,K( Eq.\,\ref{eq:Fric_WL_0K}) given in parentheses, as well as time-dependent perturbation theory (TDPT) \cite{maurer2016, Box_ab_2023}. All vibrationial lifetimes were evaluated at an electronic temperature of 300\,K. }
    \begin{tabular}{@{}ccccc@{}}
        \toprule
                                  & \multicolumn{4}{c}{\textbf{Vibrational lifetimes}}                                                  \\ \cmidrule(l){2-5} 
        \multirow{-2}{*}{ \textbf{System}} &  Brandbyge & WL   &  TDPT &  Lit.\\ \midrule
         H/Al(111)                         & 1.95      & 2.37 (2.40) & 4.54 &  \ding{55}  \\
         H/Cu(111)                         & 1.58      & 1.72 (1.75) & 1.73 &  \ding{55}  \\
         H/Pt(111)                         & 4.79      & 3.13 (3.17) & 1.13 &  0.8  \\ \bottomrule
    \end{tabular}
    \label{tab:Vib_lifetimes}
\end{table}

\section{Conclusions}
In summary, a projector operator diabatisation approach to construct Newns-Anderson model Hamiltonians
from first-principles calculations at the level of DFT was used to characterise strongly chemisorbed systems, namely hydrogen adsorbed on three fcc(111)
metal surfaces to scrutinise the applicability of the POD approach to strongly hybridised adsorbate-metal systems. We assessed the validity of this approach by reconstructing the projected density of states from the diagonalised Newns-Anderson Hamiltonian and found a strong dependence on the adsorbate basis set size. This can be rationalised by the fact that mapping the electronic structure information of the entire adsorbate-surface system onto a Newns-Anderson type Hamiltonian{---}a Hamiltonian with a single adsorbate state{---}works best if the subspace of adsorbate basis functions contains only a single basis function, which is the trivial case, as there is no reduction. By systematically increasing the subspace of the adsorbate functions from this minimal basis, i.e., a single 1$s$ orbital to a larger number, interaction of the hydrogen atom with the metal surface causes hybridisation between the adsorbate basis functions and the substrate basis functions, along with mixing among the adsorbate basis functions themselves. Hence, projecting out a single adsorbate state results in loss of information of the electronic structure, giving rise to NAH-PDOSs that differ from the PDOS acquired via diagonalising the entire block-diagonalised Hamiltonian, $\bar{\mathbf{H}}^{\text{PODGS2}}$, i.e., the larger the adsorbate basis set, the larger the deviations. Orthogonalisation of the adsorbate basis functions further shifts the lowest diabatic adsorbate state energies towards more negative values, which, for large basis sets or close adsorbate-surface distances, may lie outside of the energetic region of the metal valence DOS. This increases the electronic tunnelling times, as the chemisorption function at energies below the DOS is small. The strong position dependence of the diabatic adsorbate state at small adsorbate-surface distances reflects the strong hybridisation between the electronic states of adsorbate and surface.    The obtained chemisorption functions for the transition metal surfaces (Cu, Pt) show a pronounced energy dependence, whereas H/Al(111) provides a rather energy-independent chemisorption function. Moreover, the shift function{---}a measure for the deviation from the wideband limit approximation{---}was found to be small for H/Al(111) but pronounced on Cu and Pt. We therefore conclude that the validity of the wideband limit may even be questionable for transition metal surfaces whose Fermi-edge is dominated by $s$-bands, depending on the observable in question. Our obtained electronic tunnelling lifetimes agree with earlier computational values. Our POD-based vibrational lifetimes for the perpendicular vibrational stretch mode of hydrogen are also in qualitative agreement with TPDT-based calculations.

One aspect of our planned future endeavours is to address the strong basis-set dependence of the extracted quantities.  Currently, our approach works well when the number of basis functions for the adsorbate is small; otherwise, the orthogonalisation leads to unrealistic shifts of the adsorbate state. In this work, after testing different basis sets, we identified the Tier1-s basis set to be the best compromise between the accuracy of the DFT  ground state energy and the quality of the POD-based quantities.  This unfortunately means that, for chemisorbed systems, construction of chemisorption functions and NAHs by projection-based orthogonalisation is anything but black-box and custom basis sets may need to be developed in the future. Another interesting research pathway is to apply machine learning to the chemisorption function as a function of the adsorbate's position and energy, which would enable us to construct Newns-Anderson Hamiltonian surrogate models that automatically go beyond the wideband limit. The one-dimensional energy curves presented here could immediately be used to parametrise one-dimensional model Hamiltonians to simulate coupled electron-nuclear dynamics at surfaces.



\section*{Acknowledgement}
Funding is acknowledged from the UKRI Future Leaders Fellowship programme (MR/X023109/1), a UKRI frontier research grant (EP/X014088/1), and an MSCA postdoctoral fellowship (EP/Z001498/1). High-performance computing resources were provided via the Scientific Computing Research Technology Platform of the University of Warwick.

\section*{Supplemental Information}

\setcounter{figure}{0}
\setcounter{table}{0}
\setcounter{equation}{0}

\renewcommand{\thefigure}{S\arabic{figure}}
\renewcommand{\thetable}{S\arabic{table}}
\renewcommand{\theequation}{S\arabic{equation}}

\begin{figure}[h]
    \centering
    \includegraphics[width=3.3in]{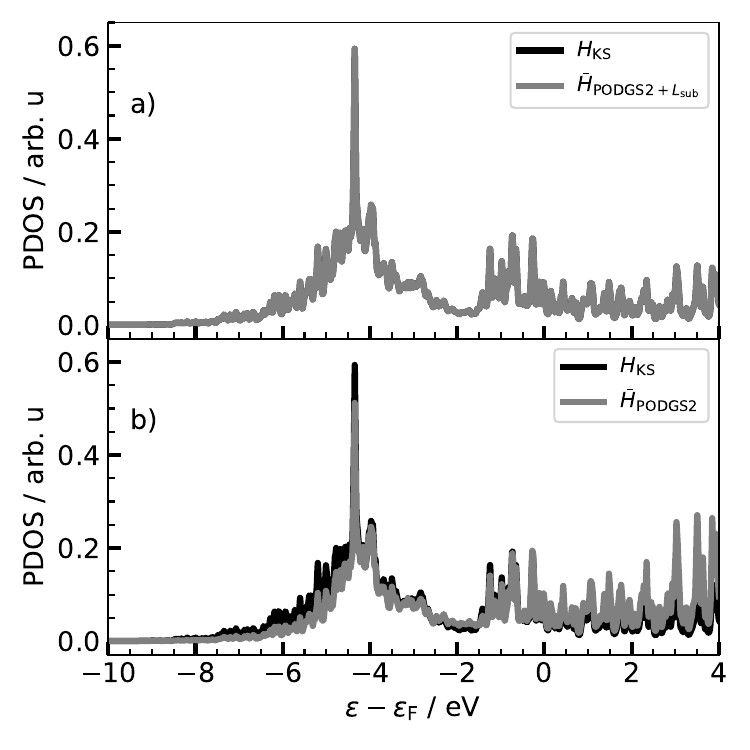}
    \caption{PDOS obtained from the Mulliken analysis of $\bar{\mathbf{H}}^\mathrm{POD2GS}$ (grey lines) and the Kohn-Sham Hamiltonian matrix $\mathbf{H}_\text{KS}$ (black lines). Panels a) and b) show the result with and without the additional L\"owdin orthogonalisation of the substrate states, respectively.}
    \label{fig:Lowdin_ss}
\end{figure}

\clearpage
\pagebreak
\newpage
\begin{figure}[h]
    \centering
    \includegraphics[width=3.3in]{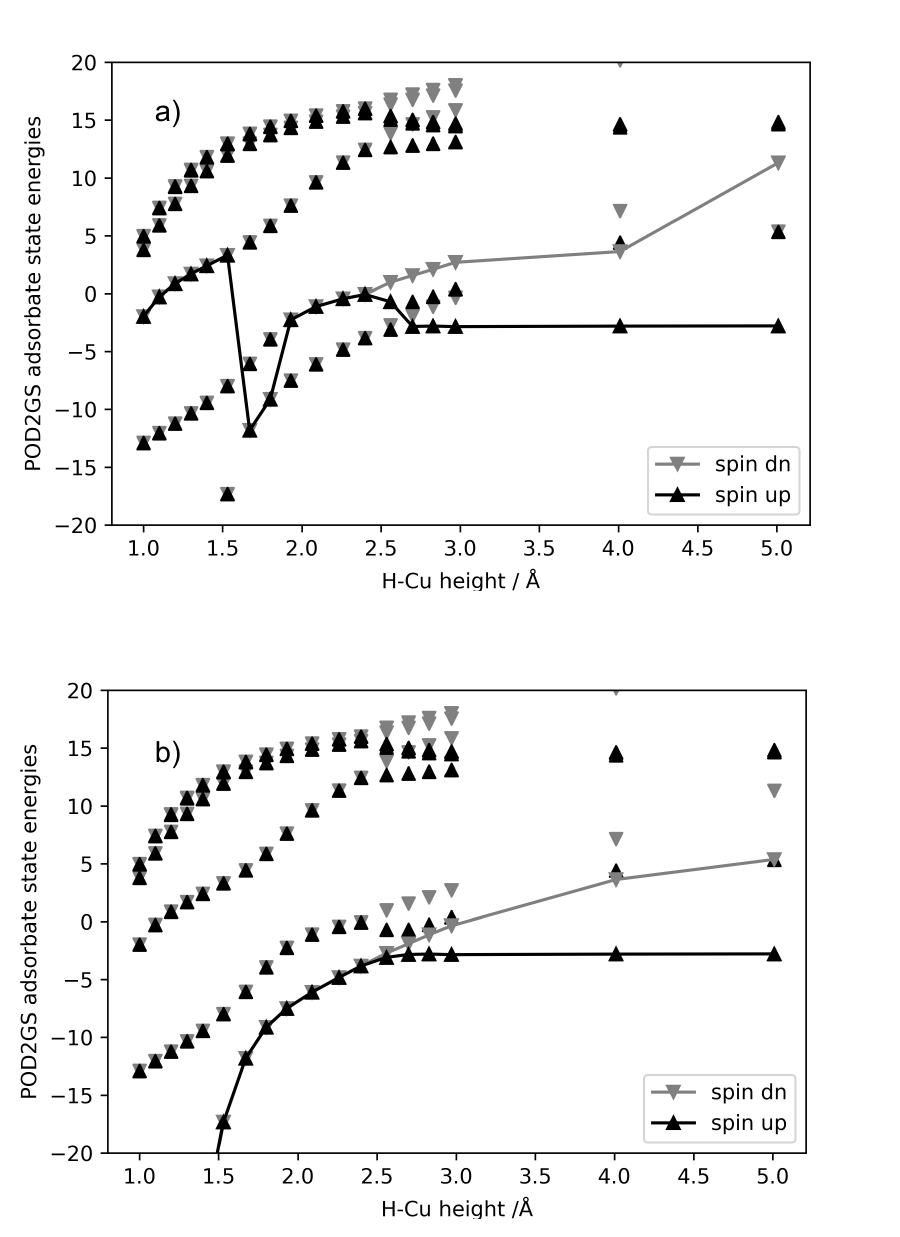}
    \caption{Diabatic adsorbate state energies for H at Cu(111) as a function of height above the surface. In panel a), the black line tracks the diabatic state with the highest overlap with the lowest diabatic state of hydrogen at a reference height of 5\,{\AA} from the surface. In panel b), the black line connects diabatic states of highest similarity to the corresponding lowest diabatic state at the preceding height step, effectively tracking the adiabatic evolution of the state across heights.  }
    \label{fig:State_following}
\end{figure}
\clearpage
\pagebreak
\newpage
\begin{figure}[h]
    \centering
    \includegraphics[width=3.3in]{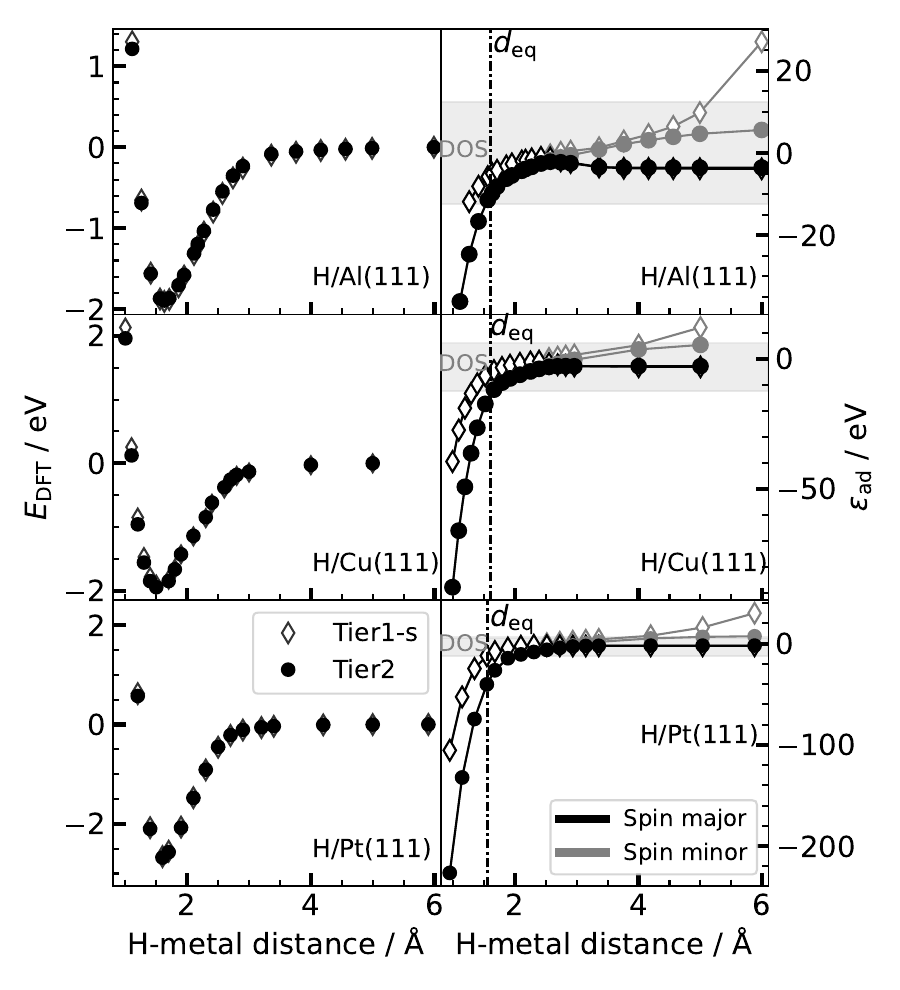}
    \caption{Adiabatic energies and diabatic adsorbate state $\varepsilon_{\text{ad}}$ calculated with the Tier2 basis sets and Tier1-s basis sets used for hydrogen. }
    \label{fig:Adiabatic_adsorbates}
\end{figure}
\clearpage
\pagebreak
\newpage

\begin{figure}[h]
    \centering
    \includegraphics[width=3.3in]{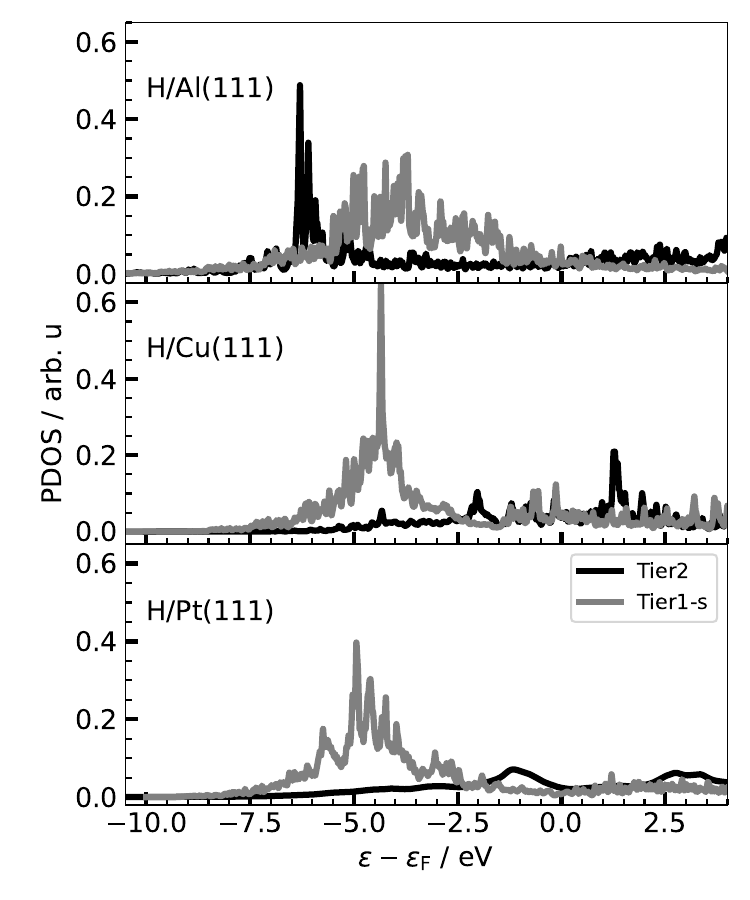}
    \caption{Projected density of states of adsorbed hydrogen at the equilibrium height, $d_\text{eq}$, computed from the NAH that was for the Tier2 basis sets and Tier1-s basis sets used for hydrogen.}
    \label{fig:PDOS_Tiers}
\end{figure}
\clearpage
\pagebreak
\newpage

\begin{table}[]
    \centering
    \begin{tabular}{l c c c c}\hline

      &      &    \multicolumn{1}{c}{Basis set type}   &  &     \\ 
 Method         & min.  & Tier1-s        & Tier2 & TDPT \\ \hline
H/Al(111)  & 0.55  & 1.95    & 0.33            &     4.54             \\
H/Cu(111) & 1.58 & 1.72    & 0.46           &     1.73         \\ 
H/Pt(111) &0.42  & 4.79    & 2.03            &   1.13           \\ \hline
    \end{tabular}
    \caption{POD-based vibrational lifetimes, $\tau_\text{vib}$, of H/Al(111), H/Cu(111), and H/Pt(111) for various NAO basis sets starting from the minimum basis set (labelled 'min' in the table) to the Tier2 basis set used in standard 'tight' settings obtained with Eq. 19 from the main manuscript. We also conducted DFT-based time-dependent perturbation theory calculations to acquire reference values for the vibrational lifetimes. The electronic temperature in all vibrational lifetime calculations was set to 300\,K. }
    \label{tab:Vibrational_lifetimes metals}
\end{table}

\begin{table}[]
    \centering
    \begin{tabular}{l c c c }\hline

      &      &    \multicolumn{1}{c}{Basis set type}   &     \\ 
 Method         & min.  & Tier1-s        & Tier2  \\ \hline
H/Al(111)  &  -4.44 &    -4.99            &  -9.81            \\
H/Cu(111) & -5.40 & -6.45    & -17.33                    \\ 
H/Pt(111) &-8.62  & -11.93   & -40.23                        \\ \hline
    \end{tabular}
    \caption{Lowest diabatic adsorbate state, $\varepsilon_\text{ad}$ in eV, of H/Al(111), H/Cu(111), and H/Pt(111) at the minimum adsorbate height, $d_\text{eq}$, for various NAO basis sets starting from the minimum basis set (labelled 'min' in the table) to the Tier2 basis set used in standard 'tight' settings. }
    \label{tab:E_ad_metals}
\end{table}




\bibliography{bibliography_paper}

\end{document}